\documentclass[thmsa]{article}
\usepackage{amsfonts}

\input{tcilatex}
\begin{document}

\title{Generalized Goldstone Theorem: Automatic Imposition of the Higgs Mechanism
and Application to Scale and Conformal Symmetry Breaking}
\author{A. Chodos\thanks{%
Present Address, American Physical Society, Washington, DC, chodos@aps.org} 
\\
Physics Department, Yale University, New Haven, CT 06520-8120 \and G.
Gallatin\thanks{%
gallatin@lucent.com} \\
Bell Labs, Lucent Technologies, \\
Murray Hill, NJ 07974-0636}
\maketitle

\begin{abstract}
Standard discussions of Goldstone's theorem based on a symmetry of the
action assume constant fields and global transformations, i.e.,
transformations which are independent of spacetime coordinates. By allowing
for arbitrary field distributions in a general representation of the
symmetry we derive a generalization of the standard Goldstone's theorem.
When applied to gauge bosons coupled to scalars with a spontaneously broken
symmetry the generalized theorem automatically imposes the Higgs mechanism,
i.e., if the expectation value of the scalar field is nonzero then the gauge
bosons must be massive. The other aspect of the Higgs mechanism, the
disappearance of the ``would be'' Goldstone boson, follows directly from the
generalized symmetry condition itself. We also use our generalized
Goldstone's theorem to analyze the case of a system in which scale and
conformal symmetries are both spontaneously broken.
\end{abstract}

\section{Introduction}

Symmetry, symmetry breaking, Goldstone bosons and the Higgs mechanism play a
very important role in modern physics. (See for example \cite{PS}\cite
{Weinberg}\cite{I+Z}\cite{Coleman}) Here we present a more general approach
to these ideas which explores the consequences of Goldstone's theorem for
space-time and gauge symmetries and shows that the Higgs mechanism is not as
``mysterious'' or ``miraculous'' as it is sometimes presented to be. We also
resolve some old questions regarding the the breaking of scale and conformal
symmetry.

We consider the physical consequences of actions which have a symmetry or a
set of symmetries. Precisely what we mean by this is discussed in detail
below. For concreteness we work out the details for the classical action of
a set of fields $\Phi _{a}$, which we denote by $S\left[ \Phi \right] .$ To
treat the quantum theory one computes the functional integral over all field
configurations (in a given function space) of $\exp \left( iS\left[ \Phi
\right] +iJ\cdot \Phi \right) $ to obtain the generating functional $Z\left[
J\right] .$ Here the dot $\,$indicates the appropriate inner product over
spacetime position, spacetime indices and internal indices. The effective
action $\Gamma \left[ \Phi \right] ,$ \cite{PS}\cite{Weinberg}\cite{I+Z}\cite
{Coleman} defined via a Legendre transformation of $\ln \left( Z\left[
J\right] \right) $, is the analog of $S\left[ \Phi \right] $ but includes
all quantum effects. That is, whereas $\delta S/\delta \Phi \left( x\right)
=0\,$is the equation of motion for the classical field configuration $\Phi
_{C}\left( x\right) $, $\delta \Gamma /\delta \Phi \left( x\right) =0$ is
the equation of motion for the vacuum expectation value of the field $%
\left\langle \Phi \left( x\right) \right\rangle .$ Indeed the $n^{th}$
functional derivative of $\Gamma \left[ \Phi \right] $ is the $n$-point
quantum Greens function. Generally, but certainly not always, $\Gamma \left[
\Phi \right] $ will have the same symmetry properties as $S\left[ \Phi
\right] $. To lowest or ``tree'' order the two actions coincide and in a
sense $\Gamma \left[ \Phi \right] $ can be thought of as just a more
complicated functional of $\Phi $ than $S\left[ \Phi \right] .$ Hence our
approach applies to both $S\left[ \Phi \right] $ and to $\Gamma \left[ \Phi
\right] $ for any given symmetry that holds for either functional. We will
generally assume that $\Phi $ is classically a commuting field or quantum
mechanically a bosonic field, but the same approach can be applied to
classical Grassmann fields or to quantum Fermionic fields yielding similar
results with, of course, the requisite care in factor ordering.

\section{General Symmetry Condition}

Consider a Lagrangian density $\mathcal{L}\left( \Phi \left( x\right)
,\partial \Phi \left( x\right) \right) $ which depends on a set of fields $%
\Phi _{a}\,$and their first derivatives, $\partial _{\mu }\Phi _{a}.$ Here $a
$ can be an internal index, a spacetime index or a combination of the two.
The action is defined by 
\begin{equation}
S\left[ \Phi \right] =\int d^{D}x\mathcal{L}\left( \Phi \left( x\right)
,\partial \Phi \left( x\right) \right)   \label{action}
\end{equation}
and is taken to be invariant, i.e., $S\left[ \Phi \right] =S\left[ \Phi
^{\prime }\right] ,$ under a continuous set of transformations of the fields
given by 
\begin{eqnarray}
\left. \Phi _{a}\left( x\right) \rightarrow \Phi _{a}^{\prime }\left(
x\right) \right.  &=&F\left( \Phi \left( x\right) ,\partial \Phi \left(
x\right) ,...,x\right)   \label{phi tran} \\
&=&\Phi _{a}\left( x\right) +\Delta _{a}\left( \Phi \left( x\right)
,\partial \Phi \left( x\right) ,...,x\right) +\cdots   \nonumber
\end{eqnarray}
where the last expression is the infinitesimal form of the transformation,
i.e, $\Delta _{a}\ll 1$. In terms of the Lagrangian this symmetry has the
form

\ 

$\int d^{D}x\mathcal{L}\left( \Phi \left( x\right) ,\partial \Phi \left(
x\right) \right) $%
\begin{eqnarray}
&=&\int d^{D}x\mathcal{L}\left( \Phi ^{\prime }\left( x\right) ,\partial
\Phi ^{\prime }\left( x\right) \right)   \label{S sym} \\
&=&\int d^{D}x\mathcal{L}\left( F\left( \Phi \left( x\right) ,\partial \Phi
\left( x\right) ,...,x\right) ,\partial F\left( \Phi \left( x\right)
,\partial \Phi \left( x\right) ,...,x\right) \right)   \nonumber
\end{eqnarray}
We have assumed the change in $\Phi _{a}$ may depend locally on $\Phi _{a}$
and possibly explicitly on $x$ as well. For ease of notation we will
abbreviate the $\Phi $ dependence of $\mathcal{L},$ $F$ and $\Delta $ as 
\begin{eqnarray}
\mathcal{L}\left( \Phi \left( x\right) ,\partial \Phi \left( x\right)
\right)  &\equiv &\mathcal{L}\left[ \Phi \left( x\right) \right] 
\label{notation} \\
F_{a}\left( \Phi \left( x\right) ,\partial \Phi \left( x\right)
,...,x\right)  &\equiv &F_{a}\left[ \Phi \left( x\right) ,x\right]  
\nonumber \\
\Delta _{a}\left( \Phi \left( x\right) ,\partial \Phi \left( x\right)
,...,x\right)  &\equiv &\Delta _{a}\left[ \Phi \left( x\right) ,x\right]  
\nonumber
\end{eqnarray}
Infinitesimally a symmetry is simply the statement that the gradient of the
action at any point in function space is perpendicular to the direction
defined by $\Delta _{a}$, i.e, 
\begin{equation}
\int d^{D}x\frac{\delta S\left[ \Phi \right] }{\delta \Phi _{a}\left(
x\right) }\Delta _{a}\left[ \Phi \left( x\right) ,x\right] =0
\label{grad S sym}
\end{equation}

The symmetry condition expressed in (\ref{grad S sym}) is not an equation
for $\Phi $ and in fact must hold for all values of $\Phi .$ Thus all the
functional derivatives of (\ref{grad S sym}) with repect to $\Phi $ must
vanish. This condition effectively assumes the equivalent of analyticity of $%
S\left[ \Phi \right] $ in function space. If the same idea is applied to $%
\Gamma \left[ \Phi \right] ,$ the effective action computed by evaluating a
functional integral, then the vanishing of the various functional
derivatives are termed ``generalized Ward-Takahashi identities''. The
various ways in which (\ref{grad S sym}) can be satisfied, which lead to
Noether's theorem and the distinction between internal and external
symmetries are discussed in the Appendix.

Taking one functional derivative of the symmetry condition (\ref{grad S sym}%
) and evaluating it at the equation of motion yields, as discussed below,
what can be seen as a generalized Goldstone theorem 
\begin{equation}
\int d^{D}x^{\prime }\left( \frac{\delta ^{2}S}{\delta \Phi _{a}\left(
x\right) \delta \Phi _{b}\left( x^{\prime }\right) }\Delta _{a}\left[ \Phi
\left( x^{\prime }\right) ,x^{\prime }\right] \right) _{\Phi _{C}}=0
\label{gen S Goldstone}
\end{equation}
Substituting (\ref{action}) and assuming locality of $\Delta _{a}$ as
expressed in (\ref{notation}) yields the following after some simple
manipulations

\begin{equation}
\begin{array}{c}
\left[ -\partial _{\alpha }\left( \frac{\partial ^{2}\mathcal{L}}{\partial
\left( \partial _{\alpha }\Phi _{a}\right) \partial \left( \partial _{\mu
}\Phi _{b}\right) }\partial _{\mu }\Delta _{b}\right) -\partial _{\alpha
}\left( \frac{\partial ^{2}\mathcal{L}}{\partial \left( \partial _{\alpha
}\Phi _{a}\right) \partial \Phi _{b}}\Delta _{b}\right) \right. \\ 
\left. +\frac{\partial ^{2}\mathcal{L}}{\partial \Phi _{a}\partial \left(
\partial _{\mu }\Phi _{b}\right) }\partial _{\mu }\Delta _{b}+\frac{\partial
^{2}\mathcal{L}}{\partial \Phi _{a}\partial \Phi _{b}}\Delta _{b}\right]
_{\Phi _{C}}=0
\end{array}
\label{gen L Goldstone}
\end{equation}
The remainder of the paper explores some of the consequences of this
equation.

\section{Goldstone's Theorem}

There are nominally two consequences of Goldstone's theorem. The primary one
is the requirement for the existence of some number of massless bosons,
called Goldstone bosons, in the theory if the symmetry is spontaneously
broken. The secondary condition is that the Goldstone bosons decouple from
the other degrees of freedom in the limit of zero momentum.

Symmetry breaking is based on the additive nature of (\ref{phi tran}) which
indicates that a given field configuration is not invariant under the
symmetry transformation. In particular it is often the case that $\Delta
_{a}\equiv 0\,$if and only if $\Phi _{a}=0$ and so for any nonzero field
configuration the transformation in (\ref{phi tran}) $\,$is inhomogeneous
and the symmetry is spontaneously broken, i.e., a nonzero field
configuration is not invariant under the symmetry transformation whereas the
zero field configuration is invariant.

We begin by reviewing the standard approach to Goldstones theorem as given
for example in the book by Peskin and Schroeder \cite{PS} This form of the
derivation proceeds by considering constant fields and field transformations
for Lagrangians of the form 
\begin{equation}
\mathcal{L}=T\left( \partial \Phi \right) -V\left( \Phi \right)
\label{stan L}
\end{equation}
Specializing to constant fields the equations of motion reduce to 
\begin{equation}
\left. \frac{\partial V}{\partial \Phi _{a}}\right| _{\Phi _{C}}=0
\end{equation}
which shows that the constant field $\Phi _{C}$ is an extremum, commonly a
minimum, of $V.$ Expanding $V$ about this miminum yields 
\begin{equation}
V\left( \Phi \right) =V\left( \Phi _{C}\right) +\frac{1}{2}\left( \Phi -\Phi
_{C}\right) _{a}\left( \Phi -\Phi _{C}\right) _{b}\left( \frac{\partial ^{2}V%
}{\partial \Phi _{a}\partial \Phi _{b}}\right) _{\Phi _{C}}+\ldots
\end{equation}
The coefficient of the quadratic term is a symmetric matrix, known as the
mass matrix, whose eigenvalues are the square of the masses of the various
fields obtained from $\left( \Phi -\Phi _{C}\right) _{a}$ after applying the
linear transformation which diagonalizes the matrix. Since the symmetry
condition is taken to hold for arbitrary field configurations it holds as
well for constant fields in which case the potential $V$ itself is
invariant, i.e., 
\begin{equation}
V\left( \Phi \right) =V\left( \Phi +\Delta \right)
\end{equation}
which implies 
\begin{equation}
\frac{\partial V}{\partial \Phi _{a}}\Delta _{a}\left( \Phi \right) =0
\end{equation}
Now differentiate with respect to $\Phi _{b}$ and evaluate the result at $%
\Phi _{C}$ to obtain 
\begin{equation}
\left( \frac{\partial ^{2}V}{\partial \Phi _{b}\partial \Phi _{a}}\right)
_{\Phi _{C}}\Delta _{a}\left( \Phi _{C}\right) =0
\end{equation}
This shows that the mass matrix has a zero eigenvalue for each linearly
independent symmetry vector $\Delta \left( \Phi _{C}\right) $ satisfying the
above equation. The number of linearly independent nonzero vectors $\Delta
\left( \Phi _{C}\right) $ is referred to as the number of broken generators, 
$N_{B}$, and hence there is one Goldstone boson for each broken generator.
Our more general result (\ref{gen S Goldstone}) or equivalently (\ref{gen L
Goldstone}) yields this same condition as well if we assume Lagrangians of
the form given in (\ref{stan L}) and set $\Phi _{C}$ to a constant field in
equation (\ref{gen S Goldstone}) or (\ref{gen L Goldstone}) which can now
clearly be seen as a generalization of the above equation.

The decoupling follows from considering a change of variables which
diagonalizes the mass matrix. This can be done by writing 
\begin{equation}
\Phi _{a}=\Phi _{a}\left( \xi _{i},\rho _{j}\right)
\end{equation}
where $i=1,...,N_{B}$ and $j=1,...,N-N_{B}$ with $\xi _{i}$ and $\rho _{j}$
defined implicitly by choosing $\Phi _{a}\left( \xi _{i},\rho _{j}\right) $
so that 
\begin{eqnarray}
\left( 
\begin{array}{ll}
\frac{\partial ^{2}V\left( \Phi \left( \xi ,\rho \right) \right) }{\partial
\xi _{i}\partial \xi _{i^{\prime }}} & \frac{\partial ^{2}V\left( \Phi
\left( \xi ,\rho \right) \right) }{\partial \xi _{i}\partial \rho
_{j^{\prime }}} \\ 
\frac{\partial ^{2}V\left( \Phi \left( \xi ,\rho \right) \right) }{\partial
\rho _{j}\partial \xi _{i^{\prime }}} & \frac{\partial ^{2}V\left( \Phi
\left( \xi ,\rho \right) \right) }{\partial \rho _{j}\partial \rho
_{j^{\prime }}}
\end{array}
\right) _{\Phi _{C}} &=&\left( 
\begin{array}{ll}
\frac{\partial \Phi _{a}}{\partial \xi _{i}}\frac{\partial ^{2}V}{\partial
\Phi _{a}\partial \Phi _{b}}\frac{\partial \Phi _{b}}{\partial \xi
_{i^{\prime }}} & \frac{\partial \Phi _{a}}{\partial \xi _{i}}\frac{\partial
^{2}V}{\partial \Phi _{a}\partial \Phi _{b}}\frac{\partial \Phi _{b}}{%
\partial \rho _{j^{\prime }}} \\ 
\frac{\partial \Phi _{a}}{\partial \rho _{j}}\frac{\partial ^{2}V}{\partial
\Phi _{a}\partial \Phi _{b}}\frac{\partial \Phi _{b}}{\partial \xi
_{i^{\prime }}} & \frac{\partial \Phi _{a}}{\partial \rho _{j}}\frac{%
\partial ^{2}V}{\partial \Phi _{a}\partial \Phi _{b}}\frac{\partial \Phi _{b}%
}{\partial \rho _{j^{\prime }}}
\end{array}
\right) _{\Phi _{C}} \\
&=&\left( 
\begin{array}{cc}
0 & 0 \\ 
0 & -\left( m_{j}\left( \Phi _{C}\right) \right) ^{2}\delta _{jj^{\prime }}
\end{array}
\right)  \nonumber
\end{eqnarray}
where there is no sum on $j\,$in the last matrix. The $\xi _{i}$ are the
Goldstone bosons (the $\xi \xi $ sector of the mass matrix vanishes by
definition) and the $\rho _{i}$ are the remaining bosons, i.e., at any given
value of $\Phi _{C}$ the $\xi _{i}$ are tangent to the symmetry directions
at that point in function space whereas the $\rho _{j}$ are perpendicular to
these directions. The symmetry transformation (\ref{grad S sym}) becomes $%
\xi _{i}\rightarrow \xi _{i}+\delta \xi _{i},$ $\rho _{j}\rightarrow \rho
_{j}$ for $\Phi _{a}\rightarrow \Phi _{a}+\Delta _{a}\left( \Phi \right) \,$%
or equivalently 
\begin{equation}
\Delta _{a}\left( \Phi \right) =\frac{\partial \Phi _{a}}{\partial \xi _{i}}%
\delta \xi _{i}
\end{equation}
Note that $\delta \xi _{i}$ must be constant for $\Delta \left( \Phi \right) 
$ constant. The symmetry condition (\ref{grad S sym}) in terms of the new
variables yields 
\begin{eqnarray}
0 &=&\frac{\delta \mathcal{L}\left[ \Phi \left( \xi ,\rho \right) \right] }{%
\delta \xi _{i}}\delta \xi _{i}  \nonumber \\
&=&\frac{\partial \mathcal{L}\left[ \Phi \left( \xi ,\rho \right) \right] }{%
\partial \xi _{i}}\delta \xi _{i} \\
&=&\frac{\partial \mathcal{L}^{\prime }\left( \xi ,\rho \right) }{\partial
\xi _{i}}\delta \xi _{i}  \nonumber
\end{eqnarray}
where the $\partial _{\mu }\delta \xi _{i}$ term vanishes since $\delta \xi
_{i}$ must be constant for the symmetry to hold. Hence $\partial \mathcal{L}%
^{\prime }/\partial \xi _{i}=0$ and $\mathcal{L}^{\prime }$ can depend only
on $\partial _{\mu }\xi _{i}$. Since $\partial _{\mu }\xi _{i}$ vanishes in
the limit of zero momentum the Goldstone bosons vanish or decouple in this
limit

\section{Automatic Higgs}

Consider a $U\left( 1\right) $ gauge model in which a complex scalar field $%
\phi =\phi _{1}+i\phi _{2}$ with $\phi _{i}$ real, $i=1,2,$ is coupled in a
locally gauge invariant way to a $U\left( 1\right) $ gauge field, $A_{\mu .}$
In this case 
\begin{eqnarray}
\Phi _{a} &\rightarrow &\left( A_{\mu },\phi _{i}\right) \\
\Delta _{a} &\rightarrow &\left( \partial _{\mu }\theta ,\varepsilon
_{ij}\phi _{j}\theta \right)  \nonumber
\end{eqnarray}
with $\theta $ an arbitrary infinitesimal scalar function of position. Note
we are working in units with the electric charge $e=1$ . Take the Lagrangian
to have the form 
\begin{eqnarray}
\mathcal{L} &=&\mathcal{L}_{A}\left( \partial _{\mu }A_{\nu }\right) +%
\mathcal{L}_{\phi }\left( \partial _{\mu }\phi _{i},\phi _{i}\right) +%
\mathcal{L}_{A\phi }\left( A_{\mu },\partial _{\mu }\phi _{i},\phi
_{i}\right) \\
&=&-\frac{1}{4}F_{\mu \nu }F^{\mu \nu }+\partial _{\mu }\phi _{i}\partial
^{\mu }\phi _{i}-V\left( \phi _{i}\phi _{i}\right) +\mathcal{L}_{A\phi
}\left( A_{\mu },\partial _{\mu }\phi _{i},\phi _{i}\right)  \nonumber
\end{eqnarray}
where $F_{\mu \nu }=\partial _{\mu }A_{\nu }-\partial _{\nu }A_{\mu }$ so
that $\mathcal{L}_{A}\,$is invariant under gauge transformations and $%
\mathcal{L}_{\phi }$ is invariant under $\phi \rightarrow e^{i\theta }\phi $
for $\theta $ constant. The extra term $\mathcal{L}_{A\phi }$ is explicitly
included to ``boost'' the symmetry from a global one to a local one. It
allows for terms of the form $A\phi \partial \phi $ and $AA\phi \phi $ which
are the lowest order terms with saturated indices. That is for arbitrary,
i.e., not necessarilly infinitesimal $\theta $ the full Lagrangian is
required to be symmetric under 
\begin{eqnarray}
A_{\mu } &\rightarrow &A_{\mu }+\partial _{\mu }\theta \\
\phi &\rightarrow &e^{i\theta }\phi  \nonumber
\end{eqnarray}
We explicitly assume the only $\partial \phi \partial \phi $ term in $%
\mathcal{L}$ is the standard kinetic energy term from $\mathcal{L}_{\phi }.$

For arbitrary infinitesimal functions $\theta $ local gauge invariance
yields the following symmetry condition for the action 
\begin{equation}
0=\int d^{D}x\frac{\delta S}{\delta A_{\beta }}\partial _{\beta }\theta
+\int d^{D}x\frac{\delta S}{\delta \phi _{j}}\varepsilon _{jk}\phi _{k}\theta
\label{higgs sym}
\end{equation}
Note that $\mathcal{L}$ is locally gauge invariant and not just $S$ thus the
remaining $\int d^{D}x$ integration in the above equation can be dropped as
discussed in previous sections. The generalized symmetry equation yields two
distinct equations since we can take derivatives with respect to $A$ and $%
\phi $

\begin{eqnarray}
0 &=&\left( \int d^{D}x\frac{\delta ^{2}S}{\delta A_{\alpha }\delta A_{\beta
}}\partial _{\beta }\theta +\int d^{D}x\frac{\delta ^{2}S}{\delta A_{\alpha
}\delta \phi _{j}}\varepsilon _{jk}\phi _{k}\theta \right) _{\Phi _{C}}
\label{higgs1} \\
&=&-\partial _{\nu }\left( \frac{\partial ^{2}\mathcal{L}}{\partial \left(
\partial _{\nu }A_{\alpha }\right) \partial \left( \partial _{\mu }A_{\beta
}\right) }\partial _{\mu }\partial _{\beta }\theta \right) +\frac{\partial
^{2}\mathcal{L}}{\partial A_{\alpha }\partial \left( \partial _{\mu
}A_{\beta }\right) }\partial _{\mu }\partial _{\beta }\theta  \nonumber \\
&&-\partial _{\nu }\left( \frac{\partial ^{2}\mathcal{L}}{\partial \left(
\partial _{\nu }A_{\alpha }\right) \partial A_{\beta }}\partial _{\beta
}\theta \right) +\frac{\partial ^{2}\mathcal{L}}{\partial A_{\alpha
}\partial A_{\beta }}\partial _{\beta }\theta  \nonumber \\
&&-\partial _{\nu }\left( \frac{\partial ^{2}\mathcal{L}}{\partial \left(
\partial _{\nu }A_{\alpha }\right) \partial \left( \partial _{\mu }\phi
_{j}\right) }\partial _{\mu }\varepsilon _{jk}\phi _{k}\theta \right) +\frac{%
\partial ^{2}\mathcal{L}}{\partial A_{\alpha }\partial \left( \partial _{\mu
}\phi _{j}\right) }\partial _{\mu }\varepsilon _{jk}\phi _{k}\theta 
\nonumber \\
&&-\partial _{\nu }\left( \frac{\partial ^{2}\mathcal{L}}{\partial \left(
\partial _{\nu }A_{\alpha }\right) \partial \phi _{j}}\varepsilon _{jk}\phi
_{k}\theta \right) +\frac{\partial ^{2}\mathcal{L}}{\partial A_{\alpha
}\partial \phi _{j}}\varepsilon _{jk}\phi _{k}\theta  \nonumber
\end{eqnarray}
and 
\begin{eqnarray}
0 &=&\left( \int d^{D}x\frac{\delta ^{2}S}{\delta \phi _{i}\delta A_{\beta }}%
\partial _{\beta }\theta +\int d^{D}x\frac{\delta ^{2}S}{\delta \phi
_{i}\delta \phi _{j}}\varepsilon _{jk}\phi _{k}\theta \right) _{\Phi _{C}}
\label{higgs2} \\
&=&-\partial _{\nu }\left( \frac{\partial ^{2}\mathcal{L}}{\partial \left(
\partial _{\nu }\phi _{i}\right) \partial \left( \partial _{\mu }A_{\beta
}\right) }\partial _{\mu }\partial _{\beta }\theta \right) +\frac{\partial
^{2}\mathcal{L}}{\partial \phi _{i}\partial \left( \partial _{\mu }A_{\beta
}\right) }\partial _{\mu }\partial _{\beta }\theta  \nonumber \\
&&-\partial _{\nu }\left( \frac{\partial ^{2}\mathcal{L}}{\partial \left(
\partial _{\nu }\phi _{i}\right) \partial A_{\beta }}\partial _{\beta
}\theta \right) +\frac{\partial ^{2}\mathcal{L}}{\partial \phi _{i}\partial
A_{\beta }}\partial _{\beta }\theta  \nonumber \\
&&-\partial _{\nu }\left( \frac{\partial ^{2}\mathcal{L}}{\partial \left(
\partial _{\nu }\phi _{i}\right) \partial \left( \partial _{\mu }\phi
_{j}\right) }\varepsilon _{jk}\partial _{\mu }\left( \phi _{k}\theta \right)
\right) +\frac{\partial ^{2}\mathcal{L}}{\partial \phi _{i}\partial \left(
\partial _{\beta }\phi _{j}\right) }\varepsilon _{jk}\partial _{\beta
}\left( \phi _{k}\theta \right)  \nonumber \\
&&-\partial _{\nu }\left( \frac{\partial ^{2}\mathcal{L}}{\partial \left(
\partial _{\nu }\phi _{i}\right) \partial \phi _{j}}\varepsilon _{jk}\phi
_{k}\theta \right) +\frac{\partial ^{2}\mathcal{L}}{\partial \phi
_{i}\partial \phi _{j}}\varepsilon _{jk}\phi _{k}\theta  \nonumber
\end{eqnarray}
To simplify notation in both equation we have implicitly assumed that the
results have been evaluated at at solution to the equations of motion.

In (\ref{higgs1}) the first term vanishes identically, the second and third
terms vanish because we have not allowed for any derivative coupling of the
gauge fields. The fifth term and seventh terms vanish since there are no $%
\phi \partial A$ or $\partial A\partial \phi \,$terms. In (\ref{higgs2}) the
first and second terms vanish. Using the fact that $\partial _{\mu }\phi
_{i}\partial ^{\mu }\phi _{i}=g^{\mu \nu }\delta _{ij}\partial _{\mu }\phi
_{i}\partial _{\nu }\phi _{j}$ is the only term quadratic in the derivatives
of $\phi $ and expanding out all the derivatives of product terms we obtain 
\begin{eqnarray}
0 &=&\left( \frac{\partial ^{2}\mathcal{L}}{\partial A_{\alpha }\partial
A_{\beta }}+\frac{\partial ^{2}\mathcal{L}}{\partial A_{\alpha }\partial
\left( \partial _{\beta }\phi _{j}\right) }\varepsilon _{jk}\phi _{k}\right)
\partial _{\beta }\theta  \label{higgs1a} \\
&&+\left( \frac{\partial ^{2}\mathcal{L}}{\partial A_{\alpha }\partial \phi
_{j}}\varepsilon _{jk}\phi _{k}+\frac{\partial ^{2}\mathcal{L}}{\partial
A_{\alpha }\partial \left( \partial _{\beta }\phi _{j}\right) }\varepsilon
_{jk}\partial _{\beta }\phi _{k}\right) \theta  \nonumber
\end{eqnarray}
and 
\begin{eqnarray}
0 &=&-\left( \frac{\partial ^{2}\mathcal{L}}{\partial \left( \partial _{\nu
}\phi _{i}\right) \partial A_{\beta }}+2g^{\nu \beta }\varepsilon _{ik}\phi
_{k}\right) \partial _{\nu }\partial _{\beta }\theta  \label{higgs2a} \\
&&+\left( 
\begin{array}{c}
\frac{\partial ^{2}\mathcal{L}}{\partial \phi _{i}\partial A_{\beta }}%
-\partial _{\nu }\left( \frac{\partial ^{2}\mathcal{L}}{\partial \left(
\partial _{\nu }\phi _{i}\right) \partial A_{\beta }}\right) -4\varepsilon
_{ik}\partial ^{\beta }\phi _{k} \\ 
+\left( \frac{\partial ^{2}\mathcal{L}}{\partial \phi _{i}\partial \left(
\partial _{\beta }\phi _{j}\right) }-\frac{\partial ^{2}\mathcal{L}}{%
\partial \left( \partial _{\beta }\phi _{i}\right) \partial \phi _{j}}%
\right) \varepsilon _{jk}\phi _{k}
\end{array}
\right) \partial _{\beta }\theta  \nonumber \\
&&+\left( 
\begin{array}{c}
-2\varepsilon _{ik}\partial ^{2}\phi _{k}+\left( \frac{\partial ^{2}\mathcal{%
L}}{\partial \phi _{i}\partial \left( \partial _{\beta }\phi _{j}\right) }-%
\frac{\partial ^{2}\mathcal{L}}{\partial \left( \partial _{\beta }\phi
_{i}\right) \partial \phi _{j}}\right) \varepsilon _{jk}\partial _{\beta
}\phi _{k} \\ 
+\left( \frac{\partial ^{2}\mathcal{L}}{\partial \phi _{i}\partial \phi _{j}}%
-\partial _{\nu }\left( \frac{\partial ^{2}\mathcal{L}}{\partial \left(
\partial _{\nu }\phi _{i}\right) \partial \phi _{j}}\right) \right)
\varepsilon _{jk}\phi _{k}
\end{array}
\right) \theta  \nonumber
\end{eqnarray}
Since $\theta $ is an arbitrary function coefficients of $\theta ,\partial
\theta ,$ and $\partial \partial \theta $ must vanish independently which
yields the following set of equations 
\begin{equation}
0=\frac{\partial ^{2}\mathcal{L}}{\partial A_{\alpha }\partial A_{\beta }}+%
\frac{\partial ^{2}\mathcal{L}}{\partial A_{\alpha }\partial \left( \partial
_{\beta }\phi _{j}\right) }\varepsilon _{jk}\phi _{k}  \label{fin1}
\end{equation}
\begin{equation}
0=\frac{\partial ^{2}\mathcal{L}}{\partial A_{\alpha }\partial \phi _{j}}%
\varepsilon _{jk}\phi _{k}+\frac{\partial ^{2}\mathcal{L}}{\partial
A_{\alpha }\partial \left( \partial _{\beta }\phi _{j}\right) }\varepsilon
_{jk}\partial _{\beta }\phi _{k}  \label{fin2}
\end{equation}
\begin{equation}
0=\frac{\partial ^{2}\mathcal{L}}{\partial \left( \partial _{\nu }\phi
_{i}\right) \partial A_{\beta }}+2g^{\nu \beta }\varepsilon _{ik}\phi _{k}
\label{fin3}
\end{equation}
\begin{equation}
0= 
\begin{array}{c}
\frac{\partial ^{2}\mathcal{L}}{\partial \phi _{i}\partial A_{\beta }}%
-\partial _{\nu }\left( \frac{\partial ^{2}\mathcal{L}}{\partial \left(
\partial _{\nu }\phi _{i}\right) \partial A_{\beta }}\right) -4\varepsilon
_{ik}\partial _{\beta }\phi _{k} \\ 
+\left( \frac{\partial ^{2}\mathcal{L}}{\partial \phi _{i}\partial \left(
\partial _{\beta }\phi _{j}\right) }-\frac{\partial ^{2}\mathcal{L}}{%
\partial \left( \partial _{\beta }\phi _{i}\right) \partial \phi _{j}}%
\right) \varepsilon _{jk}\phi _{k}
\end{array}
\label{fin4}
\end{equation}
\begin{equation}
0= 
\begin{array}{c}
-2\varepsilon _{ik}\partial ^{2}\phi _{k} \\ 
+\left( \frac{\partial ^{2}\mathcal{L}}{\partial \phi _{i}\partial \left(
\partial _{\beta }\phi _{j}\right) }-\frac{\partial ^{2}\mathcal{L}}{%
\partial \left( \partial _{\beta }\phi _{i}\right) \partial \phi _{j}}%
\right) \varepsilon _{jk}\partial _{\beta }\phi _{k} \\ 
+\left( \frac{\partial ^{2}\mathcal{L}}{\partial \phi _{i}\partial \phi _{j}}%
-\partial _{\nu }\left( \frac{\partial ^{2}\mathcal{L}}{\partial \left(
\partial _{\nu }\phi _{i}\right) \partial \phi _{j}}\right) \right)
\varepsilon _{jk}\phi _{k}
\end{array}
\label{fin5}
\end{equation}
For a constant solution to the $\phi $ equations of motion,i.e., $\phi =$
constant but $A_{\mu }$ unspecified, these equations reduce to 
\begin{equation}
0=\frac{\partial ^{2}\mathcal{L}}{\partial A_{\alpha }\partial A_{\beta }}+%
\frac{\partial ^{2}\mathcal{L}}{\partial A_{\alpha }\partial \left( \partial
_{\beta }\phi _{j}\right) }\varepsilon _{jk}\phi _{k}  \label{fin1a}
\end{equation}
\begin{equation}
0=\frac{\partial ^{2}\mathcal{L}}{\partial A_{\alpha }\partial \phi _{j}}%
\varepsilon _{jk}\phi _{k}  \label{fin2a}
\end{equation}
\begin{equation}
0=\frac{\partial ^{2}\mathcal{L}}{\partial \left( \partial _{\nu }\phi
_{i}\right) \partial A_{\beta }}+2g^{\nu \beta }\varepsilon _{ik}\phi _{k}
\label{fin3a}
\end{equation}
\begin{equation}
0= 
\begin{array}{c}
\frac{\partial ^{2}\mathcal{L}}{\partial \phi _{i}\partial A_{\beta }}%
-\partial _{\nu }\left( \frac{\partial ^{2}\mathcal{L}}{\partial \left(
\partial _{\nu }\phi _{i}\right) \partial A_{\beta }}\right) \\ 
+\left( \frac{\partial ^{2}\mathcal{L}}{\partial \phi _{i}\partial \left(
\partial _{\beta }\phi _{j}\right) }-\frac{\partial ^{2}\mathcal{L}}{%
\partial \left( \partial _{\beta }\phi _{i}\right) \partial \phi _{j}}%
\right) \varepsilon _{jk}\phi _{k}
\end{array}
\label{fin4a}
\end{equation}
\begin{equation}
0=\left( \frac{\partial ^{2}\mathcal{L}}{\partial \phi _{i}\partial \phi _{j}%
}-\partial _{\nu }\left( \frac{\partial ^{2}\mathcal{L}}{\partial \left(
\partial _{\nu }\phi _{i}\right) \partial \phi _{j}}\right) \right)
\varepsilon _{jk}\phi _{k}  \label{fin5a}
\end{equation}
Substituting (\ref{fin3a}) into (\ref{fin1a}) yields 
\[
\frac{\partial ^{2}\mathcal{L}}{\partial A_{\alpha }\partial A_{\beta }}%
=2g^{\alpha \beta }\phi _{i}\phi _{i} 
\]
and so for constant solutions to the equation, $\phi _{i},$ which are not
zero, the gauge bosons must have a nonzero mass equal to $\sqrt{2\phi
_{i}\phi _{i}}$. The sign of the gauge boson mass is correct since the
Lagrangian for massive vector bosons, the Proca Lagrangian, has the form $-%
\frac{1}{4}F^{2}+\frac{1}{2}M^{2}A^{2}.\,$Hence the first part of the Higgs
mechanism, the gauge bosons acquire a mass, is automatic and can be seen to
be simply a direct requirement of the generalized Goldstone's theorem,
equation (\ref{gen S Goldstone}) or equivalently (\ref{gen L Goldstone}).

Applying $\varepsilon _{il}\phi _{\frak{l}}$ to (\ref{fin4a}) equation and
using (\ref{fin2a}) yields 
\begin{equation}
\partial _{\nu }\left( \frac{\partial ^{2}\mathcal{L}}{\partial \left(
\partial _{\nu }\phi _{i}\right) \partial A_{\beta }}\right) \varepsilon
_{il}\phi _{\frak{l}}=0
\end{equation}
And applying $\varepsilon _{i\frak{l}}\phi _{\frak{l}}$ to (\ref{fin5a})
yields 
\begin{equation}
\varepsilon _{il}\phi _{l}\frac{\partial ^{2}\mathcal{L}}{\partial \phi
_{i}\partial \phi _{j}}\varepsilon _{jk}\phi _{k}=\varepsilon _{il}\phi
_{l}\partial _{\nu }\left( \frac{\partial ^{2}\mathcal{L}}{\partial \left(
\partial _{\nu }\phi _{i}\right) \partial \phi _{j}}\right) \varepsilon
_{jk}\phi _{k}
\end{equation}
These two equations along with (\ref{fin2a}) are automatically satisfied by
taking

\begin{equation}
\mathcal{L}_{A\phi }=2A_{\mu }\varepsilon _{ij}\phi _{i}\partial ^{\mu }\phi
_{j}+\phi _{i}\phi _{i}A_{\mu }A^{\mu }
\end{equation}
which is the standard form.

The second part of the Higgs mechanism, the disappearance of the
``would-be'' Goldstone boson, follows from the symmetry condition (\ref
{higgs sym}) itself which after changing variables to $\xi $ and $\rho $
using $\phi =\rho e^{i\xi }$ reads 
\begin{equation}
0=\frac{\partial \mathcal{L}}{\partial \left( \partial _{\mu }A_{\nu
}\right) }\partial _{\mu }\partial _{\nu }\theta +\left( \frac{\partial 
\mathcal{L}}{\partial A_{\nu }}+\frac{\partial \mathcal{L}}{\partial \left(
\partial _{\nu }\xi \right) }\right) \partial _{\nu }\theta +\left( \frac{%
\partial \mathcal{L}}{\partial \xi }\right) \theta
\end{equation}
Again, since $\theta $ is an arbitrary function, each term must vanish
separately. The first term vanishes due to the gauge invariance $\mathcal{L}%
_{A} $ since $\partial \mathcal{L}/\partial \left( \partial _{\mu }A_{\nu
}\right) =\partial \mathcal{L}_{A}/\partial \left( \partial _{\mu }A_{\nu
}\right) $. The last term demands $\partial \mathcal{L}/\partial \xi =0$ and
so $\mathcal{L}$ may depend only on derivatives of $\xi ,$ i.e., $\mathcal{L}%
\left( A,\rho ,\partial \rho ,\xi ,\partial \xi \right) \rightarrow \mathcal{%
L}\left( A,\rho ,\partial \rho ,\partial \xi \right) $. If we make the
change of variables $A_{\mu }\rightarrow B_{\mu }=A_{\mu }-\partial _{\mu
}\xi ,$ the first and third terms still vanish. the first automatically
since for $\mathcal{L}_{A}$ alone this is just a gauge tranformation and the
last still yields the condition $\partial \mathcal{L}/\partial \xi =0$. But
now the middle term can be written as 
\begin{eqnarray}
0 &=&\left( \frac{\partial \mathcal{L}\left( B,\rho ,\partial \rho ,\partial
\xi \right) }{\partial B_{\nu }}\right) \frac{\partial B_{\nu }}{\partial
A_{\mu }}+\left( \frac{\partial \mathcal{L}\left( B,\rho ,\partial \rho
,\partial \xi \right) }{\partial B_{\nu }}\right) \frac{\partial B_{\nu }}{%
\partial \left( \partial _{\mu }\xi \right) } \\
&&+\left( \frac{\partial \mathcal{L}\left( B,\rho ,\partial \rho ,\partial
\xi \right) }{\partial \left( \partial _{\nu }\xi \right) }\right)  \nonumber
\\
&=&\left( \frac{\partial \mathcal{L}\left( B,\rho ,\partial \rho ,\partial
\xi \right) }{\partial B_{\nu }}\right) \delta _{\nu }^{\mu }+\left( \frac{%
\partial \mathcal{L}\left( B,\rho ,\partial \rho ,\partial \xi \right) }{%
\partial B_{\nu }}\right) \left( -\delta _{\nu }^{\mu }\right)  \nonumber \\
&&+\left( \frac{\partial \mathcal{L}\left( B,\rho ,\partial \rho ,\partial
\xi \right) }{\partial \left( \partial _{\nu }\xi \right) }\right)  \nonumber
\\
&=&\frac{\partial \mathcal{L}\left( B,\rho ,\partial \rho ,\partial \xi
\right) }{\partial \left( \partial _{\nu }\xi \right) }  \nonumber
\end{eqnarray}
and hence $\mathcal{L}$ does not depend on $\partial \xi $ and so the
``would-be'' Goldstone boson $\xi $ has completely vanished from the model.
Effectively it has become the longitudinal component of a now massive gauge
boson.

\section{Scale and Conformal Symmetry Breaking}

It has been noted in the literature, \cite{Coleman}, see also\cite
{Polchinski}, that in theories with spontaneously broken scale and conformal
invariance, although five symmetries are broken, only one Goldstone boson
appears. A similar thing occurs for broken Lorentz invariance in a class of
three-dimensional gauge theories as discussed in \cite{Hosotani}. In this
section we use our more general treatment of Goldstone's theorem to study
this question. In particular, we see that our equation predicts only one
Goldstone mode, but also imposes four other conditions, not having to do
with particle masses, that represent the extra information contained in the
spontaneous breakdown of conformal symmetry.

We have in mind a model of the kind considered by Coleman, which contains a
scalar and a fermion field, and another scalar, the dilaton, whose role is
to implement the broken symmetry. Since we do not consider fermions in this
paper, we shall omit them here. Also, we note that the extension of the
following discussion to include more then one scalar (but still only one
dilaton) is straightforward, but to keep our notation simple we do not put
them in explicitly.

Here $\phi _{i}$ will be a doublet: $\phi _{i}=\left[ 
\begin{array}{l}
\phi \\ 
{\sigma }
\end{array}
\right] $, where $\phi $ is the ordinary scalar field and $\sigma $ is the
dilaton. Under dilations, they transform as 
\begin{equation}
\delta \phi =\phi +x^{\mu }\partial _{\mu }\phi
\end{equation}
and 
\begin{equation}
\delta \sigma ={\frac{1}{f}}+x^{\mu }\partial _{\mu }\sigma
\end{equation}
where f is a scale characterizing the symmetry breaking. Under special
conformal transformations, we have: 
\begin{eqnarray}
\delta ^{\lambda }\phi &=&(2x^{\lambda }x^{\rho }-g^{\lambda \rho
}x^{2})\partial _{\rho }\phi +2x^{\lambda }\phi  \nonumber \\
\delta ^{\lambda }\sigma &=&(2x^{\lambda }x^{\rho }-g^{\lambda \rho
}x^{2})\partial _{\rho }\sigma +\frac{2x^{\lambda }}{\text{f}}~.
\end{eqnarray}

In what follows, we shall assume that translation invariance is not broken.
Hence $\phi $ and $\sigma $ must be constants. However, one sees that the $%
\Delta $'s will not be constants. We have, in fact, 
\begin{equation}
\delta \phi =\phi ~,~~\delta \sigma ={\frac{1}{\text{f}}}~~~~\mathrm{%
(dilations)}
\end{equation}
but 
\begin{equation}
\delta ^{\lambda }\phi =2x^{\lambda }\phi ~,~~\delta ^{\lambda }\sigma =%
\frac{2x^{\lambda }}{\text{f}}~~~~\mathrm{(conformal~transfs.)}~.
\end{equation}

Noting further that the spacetime derivatives of the Lagrangian density $%
\mathcal{L}$ or of its derivatives with respect to the fields will vanish,
we obtain from (\ref{gen L Goldstone}), 
\begin{equation}
\left[ 
\begin{array}{ll}
{{\frac{\partial ^{2}\mathcal{L}}{\partial \phi ^{2}}}} & {{\frac{\partial
^{2}\mathcal{L}}{\partial \phi \partial \sigma }}} \\ 
{{\frac{\partial ^{2}\mathcal{L}}{\partial \sigma \partial \phi }}} & {{%
\frac{\partial ^{2}\mathcal{L}}{\partial \sigma ^{2}}}}
\end{array}
\right] \left[ 
\begin{array}{l}
\phi  \\ 
1/\text{f}
\end{array}
\right] =0~~~~\mathrm{(dilations)}  \label{alan dilation}
\end{equation}
and 
\begin{equation}
\left( x^{\lambda }\left[ 
\begin{array}{ll}
{{\frac{\partial ^{2}\mathcal{L}}{\partial \phi ^{2}}}} & {{\frac{\partial
^{2}\mathcal{L}}{\partial \phi \partial \sigma }}} \\ 
{{\frac{\partial ^{2}\mathcal{L}}{\partial \sigma \partial \phi }}} & {{%
\frac{\partial ^{2}\mathcal{L}}{\partial \sigma ^{2}}}}
\end{array}
\right] +\left[ 
\begin{array}{cc}
0 & \frac{\partial ^{2}\mathcal{L}}{\partial \phi \partial \left( \partial
_{\lambda }\sigma \right) }-\frac{\partial ^{2}\mathcal{L}}{\partial \sigma
\partial \left( \partial _{\lambda }\phi \right) } \\ 
\frac{\partial ^{2}\mathcal{L}}{\partial \sigma \partial \left( \partial
_{\lambda }\phi \right) }-\frac{\partial ^{2}\mathcal{L}}{\partial \phi
\partial \left( \partial _{\lambda }\sigma \right) } & 0
\end{array}
\right] \right) \left[ 
\begin{array}{l}
\phi  \\ 
1/\text{f}
\end{array}
\right] =0  \label{alan conformal}
\end{equation}
\[
~~~\mathrm{(conformal~transfs.)}
\]
In the second equation, the two terms must separately vanish, because the
first is proportional to the variable $x^{\lambda }$ and the second is not.
But the first term encodes exactly the same information as does equation (%
\ref{alan dilation}). This is the origin of the fact that dilations and
special conformal transformations give rise to the same Goldstone boson.
There is, however, the second term in (\ref{alan conformal}), which provides
an additional set of four constraints: 
\begin{equation}
{\frac{\partial ^{2}\mathcal{L}}{\partial \sigma \partial (\partial
_{\lambda }\phi )}}={\frac{\partial ^{2}\mathcal{L}}{\partial \phi \partial
(\partial _{\lambda }\sigma )}}.
\end{equation}
This is, in principle, the ``extra'' information about the Lagrangian (or
the effective action, when quantum corrections are considered) that follows
from spontaneously broken conformal symmetry.

Let us see how this works at tree level in the specific model considered by
Coleman. The Lagrange density is 
\begin{equation}
\mathcal{L}={\frac{1}{2}}\partial _{\mu }\phi \partial ^{\mu }\phi +{\frac{1%
}{2f^{2}}}\partial _{\mu }(e^{f\sigma })~\partial ^{\mu }(e^{f\sigma })-{%
\frac{\mu ^{2}}{2}}\phi ^{2}e^{2f\sigma }-{\frac{\lambda }{4!}}\phi ^{4}.
\end{equation}
The equations (\ref{alan dilation}) and (\ref{alan conformal}) imply

\begin{equation}
\left[ 
\begin{array}{ll}
1+\frac{\lambda }{2\mu ^{2}}\phi ^{2}e^{-2f\sigma } & 2\text{f}\phi \\ 
2\text{f}\phi & 2\text{f}^{2}\phi ^{2}
\end{array}
\right] \left[ 
\begin{array}{l}
\phi \\ 
1/\text{f}
\end{array}
\right] =0~,
\end{equation}
which requires $\phi =0$ and identifies $\left[ 
\begin{array}{l}
0 \\ 
1/\text{f}
\end{array}
\right] $ (i.e. the $\sigma $ particle) as the Goldstone mode. The extra
information furnished in eq. (\ref{alan conformal}) is trivial in this case,
since the relevant terms were set to zero from the beginning.

Acknowledgements. The work of A.C. was supported in part by DOE grant
\#FG02-92ER-40704.


\begin{thebibliography}{9}
\bibitem{PS}  Michael E. Peskin and Daniel V. Schroeder, ``An Introduction
to Quantum Field Theory'', Perseus Publishing, 1994.

\bibitem{Weinberg}  Steven Weinberg, ``The Quantum Theory of Fields'',
Volume 2, Cambridge University Press, 1996.

\bibitem{I+Z}  C Itzykson and J-B. Zuber, ``Quantum Field Theory'',
McGraw-Hill, 1980.

\bibitem{Coleman}  S. Coleman, ``Aspects of Symmetry'', Cambridge University
Press, 1990.

\bibitem{Polchinski}  J. Polchinski, Nuclear Physics \textbf{B303}, 226
(1988); J. Polchinski, 1992 TASI Lectures (hep-th/9210046), footnote 6.

\bibitem{Hositani}  Y. Hosotani, Proceedings DPF94, p. 1403
(QCD161,A6,1994)(hep-th/9407188)
\end{thebibliography}
\end{document}